\documentclass[a4paper]{article}
\usepackage{amsmath}
\usepackage{amsfonts}
\usepackage{amssymb}
\usepackage{graphicx}
\usepackage{tabularx}

\begin{document}
\pagenumbering{gobble}

\Large
 \begin{center}
Geomagnetic Survey Interpolation with the Machine Learning Approach\\ 

\hspace{10pt}

\large
Igor Aleshin$^{1,2}$, Kirill Kholodkov$^1$, Ivan Malygin$^1$, Roman Shevchuk$^{1,2}$, Roman Sidorov$^2$  
\hspace{10pt}

\small  
${^1)}$ Schmidt Institute of Physics of the Earth of the Russian Academy of Sciences, 123242 Moscow, Russia\\
keir@ifz.ru\\
$^{2)}$ Geophysical Center of the Russian Academy of Sciences, 119296 Moscow, Russia

\end{center}

\hspace{10pt}


\normalsize
\section{Abstract}
This paper portrays the method of UAV magnetometry survey data interpolation. The method accommodates the fact that this kind of data has a spatial distribution of the samples along a series of straight lines (similar to maritime tacks), which is a prominent characteristic of many kinds of UAV surveys. The interpolation relies on the very basic Nearest Neighbours algorithm, although augmented with a Machine Learning approach. Such an approach enables the error of less than 5 percent by intelligently adjusting the Nearest Neighbour algorithm parameters. The method was pilot tested on geomagnetic data with Borok Geomagnetic Observatory UAV aeromagnetic survey data.

\section{Introduction}
The advent of UAVs for aerial magnetometry allowed for better area coverage and faster surveys over the traditional foot-borne magnetometry\cite{prospects}. 
However, the data set obtained by geomagnetic survey using unmanned aerial vehicles (UAVs) is characterized by a high degree of spatial heterogeneity and anisotropy. It is due to the nature of the way the measurements are performed. We used a multirotor UAV with a quantum rubidium magnetometer suspended beneath. The ground speed of the UAV was approximately 3-5 m/s (6-10 kt) and the magnetic field was sampled at 10 Hz. The suspended magnetometer’s position was recorded with a GNSS receiver at the same rate. Therefore, the field sampling points were recorded at spans of about ten-to-twenty centimeters (4-8 inches) from each other.  Usually, the magnetometer survey consists of a series of straight lines. The distance between these lines is based on the characteristic spatial size of the field anomalies of interest and implemented in a flight plan for the UAV. In this work, this distance was approximately 50 meters (160 ft.) (see Fig. \ref{fig:FullData}) which is two times more than the distance between neighboring points along the line. Such spatial distribution of data aggravates the direct application of general data processing methods. The purpose of this research is to form a methodology for organizing aerial geomagnetic surveys, taking into account the widespread use of multirotor UAVs, and develop the procedures for processing the data obtained through such surveys.
Be it noted that the equivalent situation occurs during the cross-borehole electromagnetic imaging. In our paper  \cite{interwell} it was shown that the problem can be solved with a scale transformation of one of the axis. Similarly, we used the method for processing geomagnetic survey results in this paper. As a test range, we used UAV geomagnetic measurements in the vicinity of the Borok Geomagnetic Observatory. The first section provides a brief description of the measurement site, the equipment used, and the survey arrangement.
\section{Survey}
\begin{figure}[h]
\includegraphics[width=1.0\textwidth]{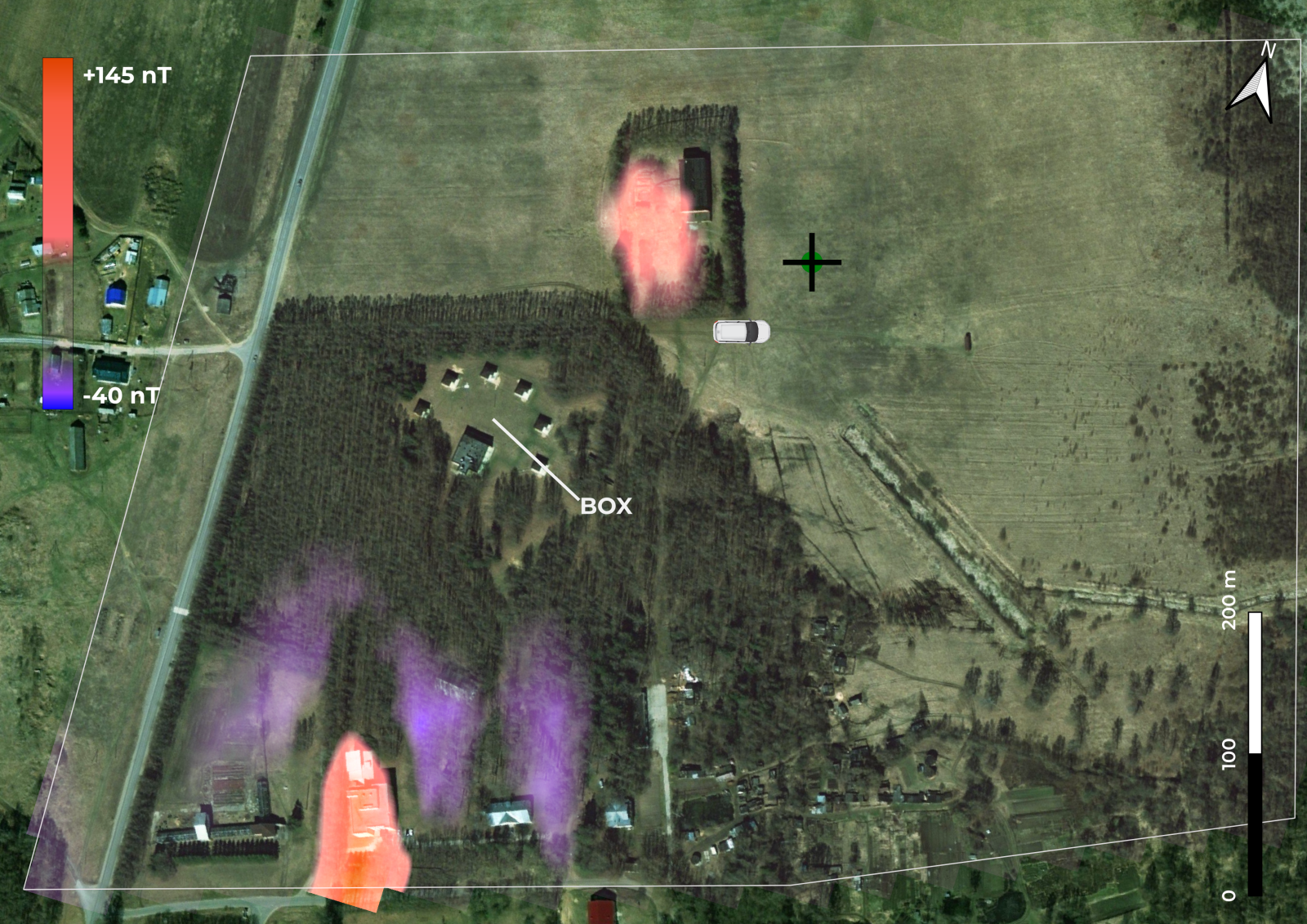}
\caption{Magnetic anomaly map superposed over the Microsoft Bing satellite imagery. The Geomagnetic Observatory “Borok” is marked with the label BOX (which is the IAGA code of the site). The car symbol shows the position of the motor vehicle with UAV control equipment. Cross depicts the place where take-off, landing, and tethering procedures were performed.}
\centering
\label{fig:Map}
\end{figure}
The measurements were carried out above the land directly adjacent to the Borok Geomagnetic Observatory \cite{borok} (see fig. \ref{fig:Map}), which participates in the international INTERMAGNET \cite{geomagneticGvishiani}, \cite{rusmagnet} network. The location of the observatory is characterized by the absence of significant sources of electromagnetic disturbances. However, this does not limit the presence of magnetic anomalies caused by nearby residential and commercial buildings. This circumstance makes the place perfect for practicing the technique of aeromagnetic survey with UAVs.  In the future, the obtained results can be partially compared with ground-based measurements, as well as used for a complete assessment of the magnetic situation around the observatory.
Due to the small area of possible anomalies, it is convenient to change geodetic latitude $\varphi$ and  longitude $\lambda$ to local cartesian coordinates centered in the middle of the region ($\phi_C$,$\lambda_C$)
$$
\varphi_C = (\varphi_{min} + \varphi_{max})/2,\quad\lambda_C=(\lambda_{min}+\lambda_{max})/2
$$
where $\varphi_{min}$, $\varphi_{max}$, $\lambda_{min}$, $\lambda_{max}$ --— minimum and maximum values of latitude and longitude, appropriately.  To do this, we will perform an orthographic projection of the points. Due to the small size of the section, we can use approximate formulas:
$$
x = (\pi R_E/180)(\lambda - \lambda_C) \cos{\varphi_C},\quad y = (\pi R_E/180)(\varphi-\varphi_C),
$$
$$
|\varphi - \varphi_C | << 1,\quad 
 |\lambda - \lambda_C | << 1.
$$
\begin{figure}[h]
\includegraphics[width=1.0\textwidth]{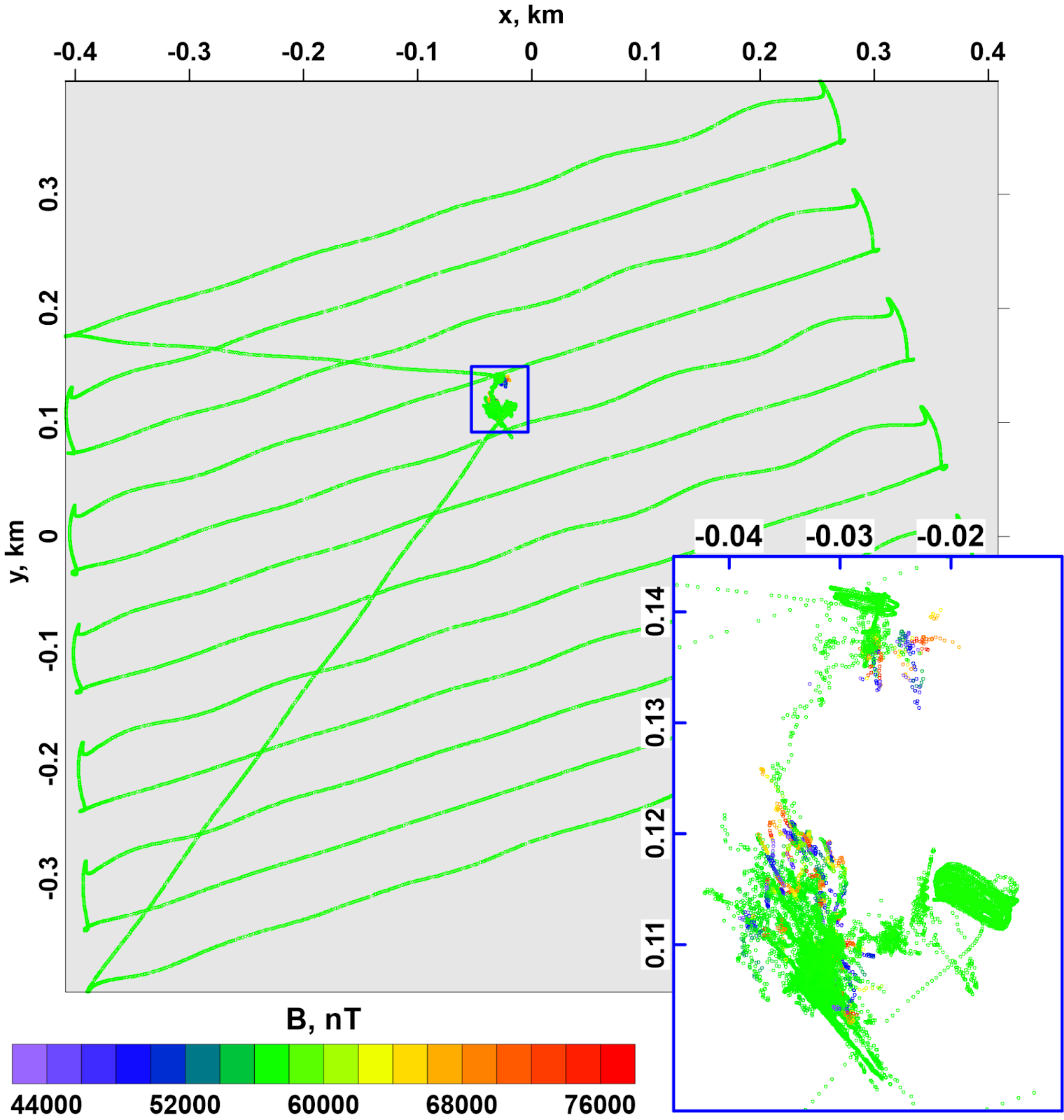}
\caption{The color-coded map of samples of the geomagnetic field. The detached map shows the detailed sampling of near the car and where take-off, landing, and tethering procedures were performed.}
\centering
\label{fig:FullData}
\end{figure}
In fig. \ref{fig:FullData} depicts the points where the geomagnetic field induction modulus was sampled. The initial values obtained by
the quantum magnetometer were adjusted to take into account the rapid temporal variations of the main magnetic field. In addition, to improve the accuracy of measurements during post-processing, the spatial position of the magnetometer was refined. These augmentations were made possible by additional equipment on-site: the proton magnetometer GEM GSM-19 and the  Javad Positioning Systems Alpha 2 GNSS receiver with a MarAnt antenna. The takeoff, landing, and magnetometer tethering position of the UAV was located near the center of the region of this research. Also, right next was a motor vehicle with UAV flight control equipment that introduces significant distortions into the measured field. In fig. F, upper panel (A), the motor vehicle introduces distortion in the nearby area seen on the magnetic field modulus as peaks of the order of magnitude higher than average values. This particular area incorporates a big part of the low-quality garbled various altitude sample points because taking-off, tethering, and landing are time-consuming and the timing of these procedures is on par with the timing of the whole survey.  Due to this, the points located in the vicinity were exempted from further processing.
\begin{figure}[h]
\includegraphics[width=1.0\textwidth]{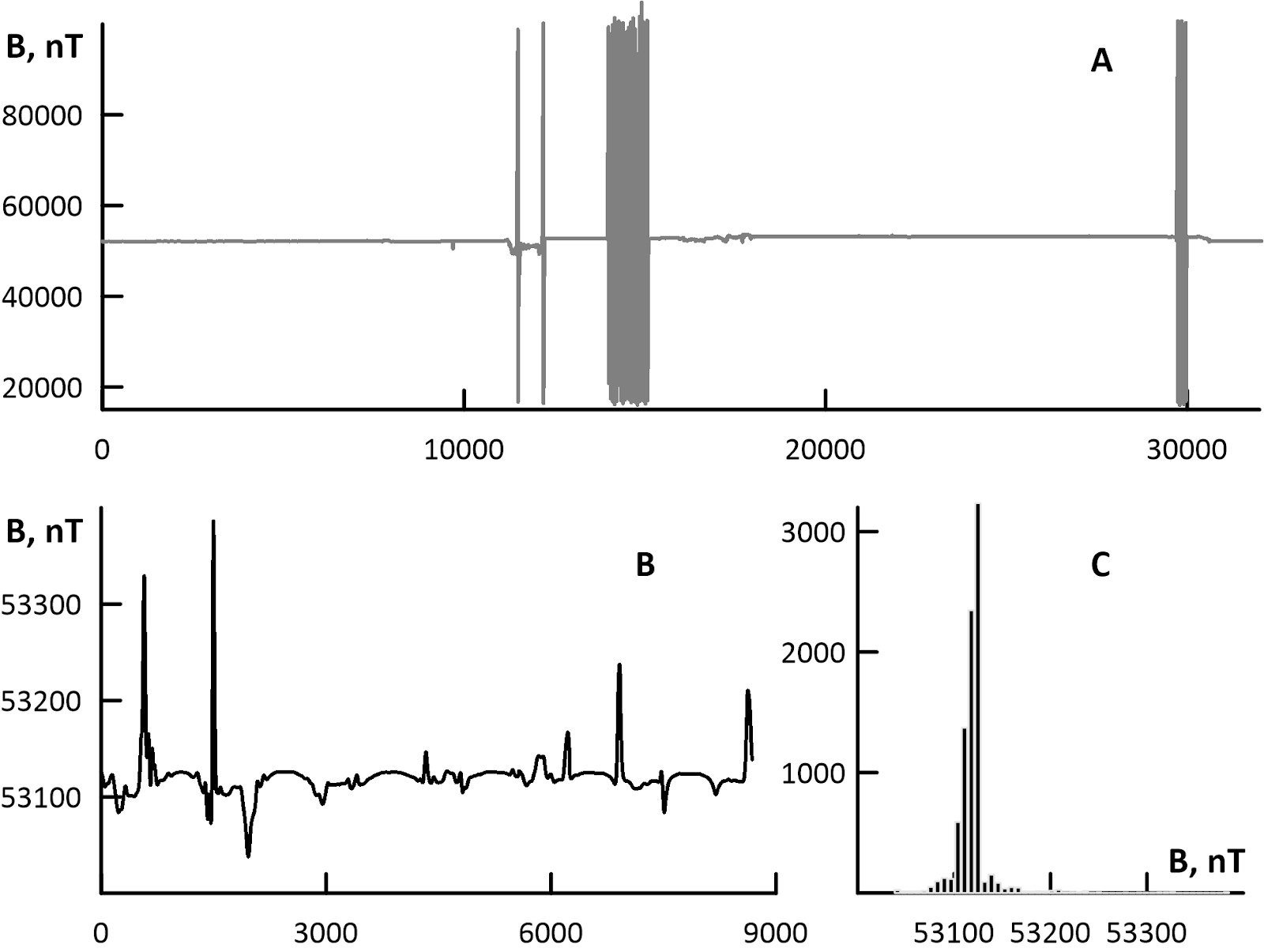}
\caption{Panel A shows the original absolute magnetic field value. Panel B shows the magnetic field with anomalies cause by take-off, landing, and tethering procedures and the car with control equipment excluded. C shows the histogram of the measured magnetic field.}
\centering
\label{fig:F}
\end{figure}
The removal of noisy records led to an almost threefold reduction in the number of source points, from 32063 to 8683. The average measured value of the magnetic field modulus essentially coincides with the median \emph{id est} $B_0\approx53120$ nT. Finally, we factored out the magnetic field variations with data from the variometer, located in the same area.
\section{Geomagnetic Field Anomaly Map}
\begin{figure}[h]
\includegraphics[width=1.0\textwidth]{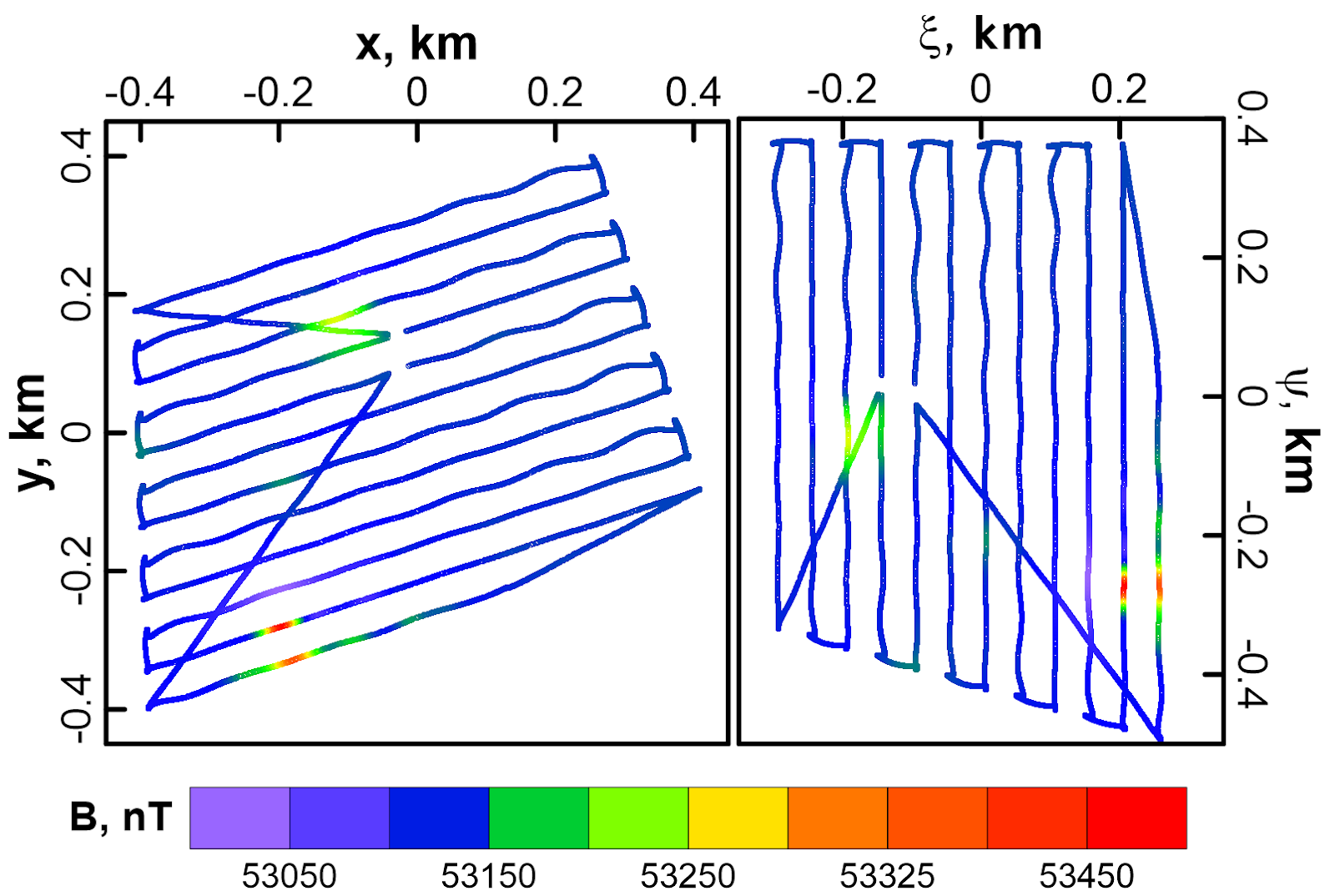}
\caption{The magnetic field on-site with extrinsic anomalies removed. A rightward panel shows the same data rotated by $\alpha \approx 72\deg$ around the coordinate origin.}
\centering
\label{fig:Cut}
\end{figure}

\begin{figure}[h]
\includegraphics[width=1.0\textwidth]{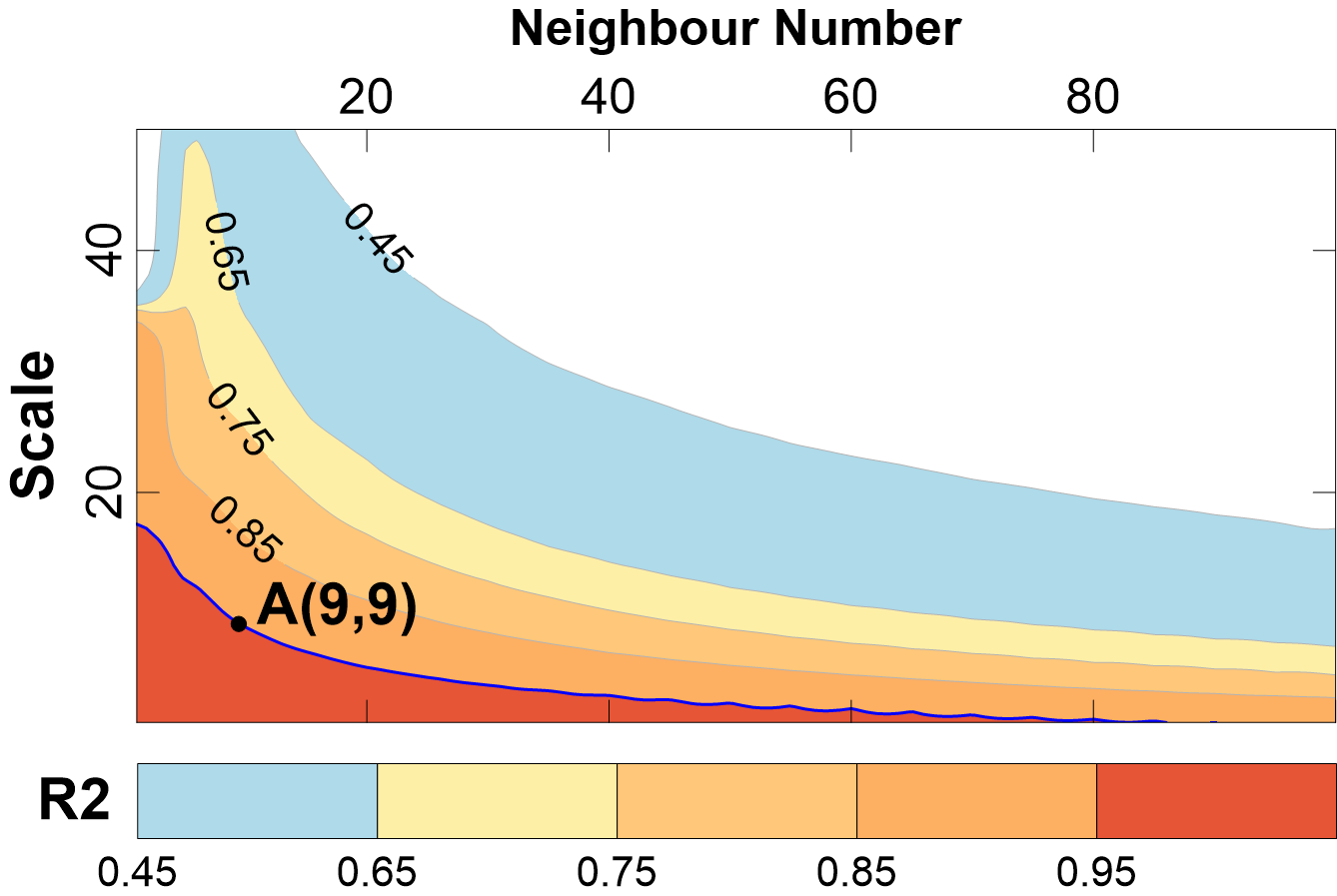}
\caption{Quality functional $R2$ computed with cross-validation. The blue contour corresponds to  $R2=0.9$; Picked hyperparameter values are denoted with label A.}
\centering
\label{fig:R2}
\end{figure}

\begin{figure}[h]
\includegraphics[width=1.0\textwidth]{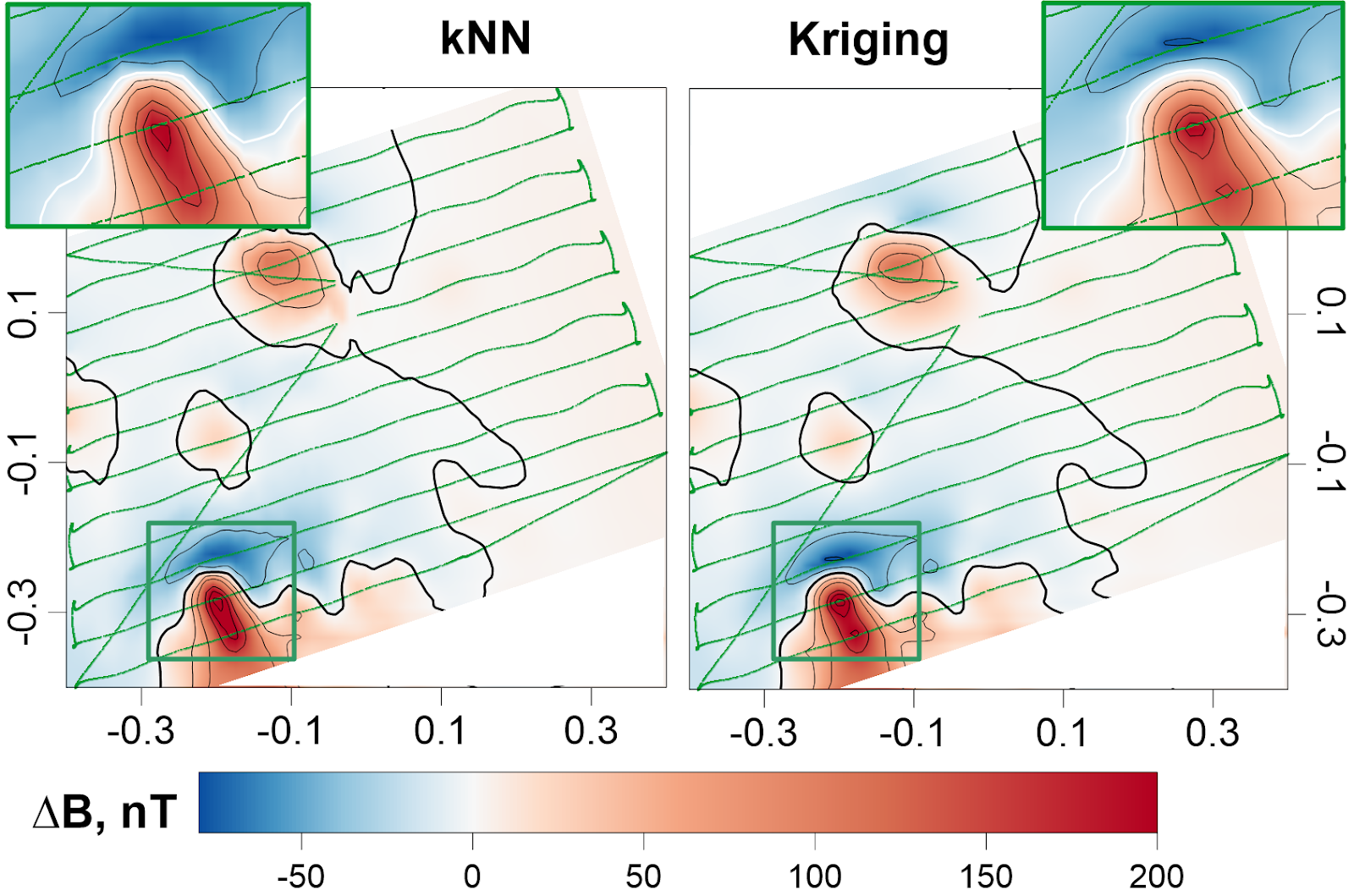}
\caption{Magnetic field variances $\delta B(x,y)$ --- the difference between the measured magnetic field and the \emph{in situ} stationary variometer. The map leftward is constructed with augmented kNN from this work. A rightward map was constructed with kriging. A thick contour denotes zero deviation.}
\centering
\label{fig:Fit}
\end{figure}

As previously mentioned in the Introduction, the distribution of the resulting data set ${b_i(x_i,y_i)}$ is strongly inhomogeneous and anisotropic. Similarly to the case of earlier research \cite{interwell}, we will use the approach described there.
To do this, we will switch to a coordinate system where the the ordinate axis is parallel to the long lines of the flight path (see Fig. \ref{fig:Cut}).
To calculate the $B(x,y)$-value at an arbitrary point $(x,y)$ with the original sparse dataset ${b_i(x_i,y_i)}$ we will use the nearest neighbors algorithm. 
\begin{equation}
    B(x,y;K,s) = 1/w\sum_{k=1}^{K} w_k(x,y|x_i,y_i;K,s),
    \label{eqn:kNN}
\end{equation}
\begin{equation}
    w_k(x,y;x_i,y_i;K,s)\sim\sqrt{(x-x_i)^2/s^2-(y-y_i)^2},\;w=\sum_{k=1}^{K}w_k
    \label{eqn:Scale}
\end{equation}
Where $K$ — the number of nearest neighbors calculated by distance $w_k$.
Scale parameters and the number of nearest neighbors $K$ can be determined using the cross-validation technique.  The initial data is split into $M$ groups, each is used to calculate the quality functional one at a time. We used the coefficient of determination as quality functional. The coefficient of determination is the proportion of the variance of the variable explained by the model. It is defined by these formulae:
$$
R2(K,s)=(1/M)\sum_{m=1}^{M}\eta^{(m)}(K,s)
$$
$$
\eta^{(m)}(K,s)=1-\sum_{i=1}^{N/M}(b^{(m)}_i(x_i,y_i)-B(x_i,y_i;K,s)^2/\sum_{i=1}^{N/M}(b^{(m)}_i(x_i,y_i)-\bar{b}^{(m)}(K,s))^2
$$
$$
\bar{b}^{(m)}(K,s)=(M/N)\sum_{i=1}^{N/M}b^{m}_i(x_i,y_i)
$$
To determine the optimal values of hyperparameters, the quality functional values $R2(K, s)$ were calculated for all combinations of integer values $1\le K \le 100$ and $1 \le s \le 50$.
The result is shown in fig. \ref{fig:R2}. It shows that the nearest neighbor model (eqn. \ref{eqn:kNN}) with modified distance (eqn. \ref{eqn:Scale}) accounts for the data very well. If $R2(K, s) \sim 1$ the interpolation error is minimal and nears zero, such values cannot be used for real. Instead, in this work, we used $R2(K, s)=0.95$. In fig. \ref{fig:R2} this value corresponds to the blue contour. Every point of this contour determines the matching K and s. In our case, we utilized the multiparametric optimization (like in \cite{hyperp}) and picked a point closest to the coordinate origin $K = s = 9$. Generally, any other point of the contour provides the visually same result.
Figure \ref{fig:Fit} left shows a map of the magnetic induction deviation from its average constructed with the data and method displayed above. Throughout most of the territory, there are no fields close to the average value $B_0$. For comparison on fig. \ref{fig:Fit}  right we constructed the same map but with kriging. The two methods appear to show similar results. However, the differences could be found in anomalous areas. Insets show, that the augmented $k$NN method performs better at avoiding of peak splitting caused by a linear layout of the samples. There are two noticeable anomalies in the northern and southern parts of the site with amplitudes of 50 nT and 150 nT, respectively. 
To explain their nature, we superimposed the resulting map on satellite imagery of the site (Fig. \ref{fig:Map}). The figure shows that the source of both inhomogeneities are artificial structures. The anomaly to the north is caused by reinforced concrete structures of sewage treatment plants. The cause of the anomaly in the southern part is a heavy rock testing equipment installed in one of the buildings. It is important to note that despite the proximity of both sources of the anomalous field to the observatory buildings, the geomagnetic field in their vicinity remains undisturbed.
\section{Conclusion}
The conducted research has shown that an aeromagnetic survey using UAVs supplemented by modern data processing methods is an effective tool for studying the geomagnetic field. In particular, the aforesaid method makes it possible to identify the main anomalies of the geomagnetic despite the significant heterogeneity and anisotropy of the distribution of the measured values. Note that the degree of heterogeneity and anisotropy of the data can be reduced by changing the flight pattern. To fight this, one needs to perform an additional series of flyovers perpendicular to the original ones. Perhaps such a scheme will be relevant for a complex configuration of geomagnetic anomalies. However, this will at least double the amount of fieldwork, even though the simplified scheme in principle allows you to identify the basic structure of the geomagnetic field. Finally, the above analysis once again demonstrates the advantages of the aerial survey compared to ground measurements: in addition to increasing productivity, the use of UAVs allows you to measure the magnitude of the magnetic induction vector directly over large artificial objects, access to which is difficult.
\section{Acknowledgments and Funding}
The authors sincerely thank the staff of the Borok Geophysical Observatory and personally Director S.~V.~Anisimov for their help in organizing and conducting geomagnetic measurements on the territory of the observatory.

The work was carried out within the framework of the state assignment of the Schmidt Institute of Physics of the Earth of the Russian Academy of Sciences (IPE RAS) and Geophysical Center of the Russian Academy of Sciences (GC RAS), approved by the Ministry of Education and Science of the Russian Federation. The collaboration between the institutes was enabled by the separate agreement on joint scientific activities between the IPE RAS and the GC RAS.

During the research, the equipment of IPE RAS and GC RAS was used.


\begin{thebibliography}{3}
\bibitem{interwell} 
Aleshin, I.M., Malygin, I. V.
\textit{Machine learning approach to inter-well radio wave survey data imaging}. 
Russian Journal of Earth Sciences. 2019. Num.~3. Vol.~19. P.~ES30031—ES30036. DOI:10.2205/2019ES000664


\bibitem{hyperp} 
Chiu, P-W., Naim, A. M., Lewis, K. E., Bloebaum, C. L.
\textit{The hyper-radial visualisation method for multi-attribute decision-making under uncertainty} International Journal of Product Development. 2009. Num.~1-3. Vol.~9. P.~4—31.

\bibitem{prospects}
Aleshin I.M., Soloviev A.A., Aleshin M.I., Sidorov R.V., Solovieva E.N., K I. Kholodkov (2020)
\textit{Prospects of Using Unmanned Aerial Vehicles in Geomagnetic Surveys} Seismic Instruments, Vol. 56, No. 5, pp. 522–530, doi:10.3103/S0747923920050059

\bibitem{borok}
Chulliat, A., Anisimov, S. V. 
\textit{The Borok INTERMAGNET magnetic observatory}
Russian Journal of Earth Sciences. 2008. Num.~3. Vol.~10. P.~ES3003. DOI:10.2205/2007ES000238

\bibitem{geomagneticGvishiani}
Gvishiani A., Soloviev A.
\textit{Observations, Modeling and Systems Analysis in Geomagnetic Data Interpretation}
Springer International Publishing. 2020. 311~p. DOI:10.1007/978-3-030-58969-1 

\bibitem{rusmagnet}
Khomutov, S. Y.
\textit{International project INTERMAGNET and magnetic observatories of Russia: cooperation and progress}
E3S Web Conf. 62 02008 (2018) DOI:10.1051/e3sconf/20186202008
\end{thebibliography}
\end{document}